\documentclass[twocolumn]{aastex62}

\usepackage{bm}
\usepackage{amsmath}
\usepackage{multirow}

\received{}
\revised{}
\accepted{}

\submitjournal{The Astrophysical Journal}

\shorttitle{Shocks and Energy Dissipation in MHD Turbulence}
\shortauthors{Park and Ryu}

\begin{document}

\title{Shock Waves and Energy Dissipation in Magnetohydrodynamic Turbulence}

\author{Junseong Park}
\author[0000-0002-5455-2957]{Dongsu Ryu}
\affil{Department of Physics, School of Natural Sciences, UNIST, Ulsan 44919, Korea}

\correspondingauthor{Dongsu Ryu}
\email{ryu@sirius.unist.ac.kr}

\begin{abstract}


Shock waves play an important role in turbulent astrophysical media by compressing the gas and dissipating the turbulent energy into the thermal energy. We here study shocks in magnetohydrodynamic turbulence using high-resolution simulations. Turbulent Mach numbers of $\mathcal{M}_\text{turb}=0.5-7$ and initial magnetic fields of plasma beta $\beta_0=0.1 - 10$ are considered, targeting turbulences in interstellar and intracluster media. Specifically, we present the statistics of fast and slow shocks, such as the distribution of shock Mach numbers ($M_\text{s}$) and the energy dissipation at shocks, based on refined methodologies for their quantifications. While most shocks form with low $M_\text{s}$, strong shocks follow exponentially decreasing distributions of $M_\text{s}$. More shocks appear for larger $\mathcal{M}_\text{turb}$ and larger $\beta_0$. Fast shock populations dominate over slow shocks if $\beta_0\gg1$, but substantial populations of slow shocks develop in the cases of $\beta\lesssim1$, i.e., strong background fields. The shock dissipation of turbulent energy occurs preferentially at fast shocks with $M_\text{s}\lesssim$ a few to several, and the dissipation at strong shocks shows exponentially decreasing functions of $M_\text{s}$. The energy dissipation at shocks, normalized to the energy injection, $\epsilon_\text{shock}/\epsilon_\text{inj}$, is estimated to be in the range of $\sim0.1-0.5$, except for the case of $\mathcal{M}_\text{turb}=0.5$ and $\beta_0=0.1$ where the shock dissipation is negligible. The fraction decreases with $\mathcal{M}_\text{turb}$; it is close to $\sim0.4-0.6$ for $\mathcal{M}_\text{turb}=0.5$, while it is $\sim0.1-0.25$ for $\mathcal{M}_\text{turb}=7$. The rest of the turbulent energy is expected to dissipate through the turbulent cascade. Our work will add insights into the interpretations of physical processes in turbulent interstellar and intracluster media.

\end{abstract}


\section{Introduction}\label{sec1}

Turbulence is present in a variety of astrophysical media, including the interstellar medium (ISM) \citep[see, e.g.,][]{Elmegreen2004,MacLow2004,McKee2007} and the intracluster medium (ICM) \citep[see, e.g.,][]{Schuecker2004,chur12,Hitomi2016,Vazza2017}. Unlike terrestrial turbulence, astrophysical turbulence is mostly transonic/supersonic and in all the cases magnetized. The properties of flows change when the average flow speed is comparable to or exceed the sound speed in the medium; high speeds usually mean that the flows are compressible. Magnetic fields permeated in turbulence are stretched and amplified by flow motions, and in turn, exert tension and pressure to flow motions altering flow properties; these processes lead to magnetohydrodynamics (MHDs). Therefore, astrophysical turbulence is by nature {\it compressible MHD turbulence} {\citep[see, e.g.,][]{Biskamp2003}}.

Observations suggest that the turbulence in the ISM is transonic or supersonic with different turbulent Mach numbers, $\mathcal{M}_\text{turb}$, in different phases. In the warm ionized medium (WIM), for instance, the width of H$\alpha$ line is in the range of tens km s$^{-1}$ \citep{tufte1999}, suggesting that $\mathcal{M}_\text{turb}$ is of order unity. For the cold neutral medium (CNM), HI observations revealed the internal gas motions corresponding to $\mathcal{M}_\text{turb} \sim$ a few \citep{Heiles2003}. Molecular clouds (MCs) are characterized with nonthermal linewidths due to highly supersonic motions of typically $\mathcal{M}_\text{turb} \gtrsim 10$ \citep[see, e.g.,][]{Larson1981}. The strength of magnetic fields in the ISM also varies with location. The strength in the Solar neighborhood, for instance, was estimated to be $B \sim 6 \mu$G, with a large-scale component of $B_\text{regular} \sim 2 - 3 \mu$G and a random component of $B_\text{random} \sim 3 - 4 \mu$G \citep{haverkorn2015}, and this corresponds to the plasma beta, the ratio of thermal to magnetic pressures, of $\beta \lesssim 1$ in the WIM. In MCs, magnetic fields are much stronger, typically of order $\sim$ mG, corresponding to $\beta \sim 0.01 - 0.1$ \citep[see, e.g.,][]{crutcher2012}.

The turbulence in the ICM is driven by ongoing accretion and mergers as well as galactic winds and feedbacks from active galactic nuclei (AGNs) \citep[see, e.g.,][]{bj14}. Simulations of cosmic structure formation have shown that the turbulence is mildly transonic with $\mathcal{M}_\text{turb} \sim 1/2$ \citep[see, e.g.,][]{ryu08,Vazza2017}, and X-ray observations support it \citep{Schuecker2004,chur12}. Observations of Faraday rotation measures and diffuse synchrotron emissions from radio halos and relics indicate the presence of $\mu$G-level magnetic fields over the whole volume of galaxy clusters \citep[see, e.g.,][]{ct02,gf04}. With these fields, the plasma beta of the ICM is estimated to be $\beta \sim 10 - 100$.

Shock waves commonly develop in astrophysical turbulences. In MCs, shocks are often responsible for the formation of filaments \citep[see, e.g.,][]{federrath2016}, and star formation appears to be linked to it as high density regions are the sites of pre-stellar cores \citep[see, e.g.,][]{MacLow2004}. Shocks are also believed to be a mechanism for driving chemical evolution in the ISM; {for instance, \citet{Pety2000} invoked turbulent dissipation in shocks or vortices to explain chemical anomalies in diffuse MCs, and \citet{Pon2012} and \citet{LehWar2016} used shocks driven by turbulence to model anomalous emission from high-J CO lines in giant MCs}. In the ICM, shocks induced by turbulent flow motions are weak, but yet they play an important role in the evolution of vorticity \citep{Porter2015} and significantly contribute to the gas heating \citep[see, e.g.,][]{Ryu2003}. Thus, describing the statistics of shocks in compressible MHD turbulence would be necessary to understand physical processes in astrophysical media.

The statistics of shocks were previously investigated, {mostly through simulations of astrophysical turbulence}. \citet{Smith2000a,Smith2000b}, for instance, presented the probability distribution function (PDF) of $M_\text{jump} \equiv v_\text{jump}/c_\text{s}$, where $v_\text{jump}$ is the velocity jump across flow-converging regions and $c_\text{s}$ is the sound speed, in decaying and driven MHD turbulences. More recently, \citet{Porter2015} and \citet{Lehmann2016} identified ``shocked'' grid zones, i.e., the grid zones through which shocks pass, in simulations, and presented the PDF of the shock Mach number, $M_\text{s} \equiv v_\text{s}/c_\text{s}$, where $v_\text{s}$ is the shock speed. While \citet{Porter2015} considered the turbulence of $\mathcal{M}_\text{turb} \sim 1/2$ in the ICM plasma with $\beta \gg 1$, \citet{Lehmann2016} considered $\mathcal{M}_\text{turb} \sim 9$ in {$\beta \lesssim 0.1$}, targeting the turbulence in MCs. Especially, \citet{Lehmann2016} divided shocks into the fast and slow populations, and separately presented their statistics. Both works showed that while weak shocks are common, strong shocks with $M_\text{s}\gg\mathcal{M}_\text{turb}$ follow PDFs exponentially decreasing with $M_\text{s}$. {\citet{Lesaffre2013}, on the other hand, attempted to study shocks in observation, by apply PDFs of $v_\text{s}$ from different shock models to interpret molecular and atomic lines in Stephan’s Quintet and also in the diffuse ISM toward Chamaeleon. They argued that in both Stephan’s Quintet and Chamaeleon, shocks of low and moderate $v_\text{s}$ are important in shaping line emissions from interstellar gas.}

In turbulent media, the turbulent energy cascades down to small scales and dissipates into heat at dissipation scales. However, {a fraction of the turbulent energy can directly dissipate at shocks}. To quantify it, often the dissipation timescale, $t_\text{diss} \sim E_\text{turb}/{\dot E}$, is compared to the flow crossing timescale, $t_\text{cross} \sim L_\text{inj}/v_\text{turb}$, expecting that while $t_\text{diss}$ counts for both the cascade and shock dissipations, while $t_\text{cross}$ does only the dissipation through turbulent cascade. Here, $E_\text{turb}$ and $v_\text{turb}$ are the turbulent energy and flow speed, respectively, ${\dot E}$ is the energy dissipation rate, and $L_\text{inj}$ is the injection scale at which the turbulence is driven. \citet{Stone1998}, for instance, estimated $t_\text{diss}$ in MHD turbulence simulations, assuming that ${\dot E}$ balances the energy injection rate in driven cases. They obtained {$t_\text{diss}/t_\text{cross} \sim 0.46 - 0.69$ for supersonic turbulences with $\beta = 0.01 - 1$, when $t_\text{diss}$ was defined with the kinetic energy}. They got a little smaller ratios for decaying turbulences. {Also, there have been attempts to estimate turbulent dissipation in observation, for instance, by comparing the observations of CO rotational transitions to detailed shock models;} \citet{Pon2014} estimated $t_\text{diss}/t_\text{cross} \sim 1/3$ for turbulent regions in the Perseus MC, and \citet{Larson2015} estimated $t_\text{diss}/t_\text{cross} \sim 0.65$ or 0.94, depending on the adopted shock model, for the Taurus MC.

There have been trials to directly estimate the amount of the energy dissipated at shocks. \citet{Lehmann2016}, for instance, calculated the kinetic energy flux, $(1/2)\rho_1v_s^3$ ($\rho_1$ is the preshock density), through shock surfaces, anticipating that it represents the shock dissipation. They estimated that for the turbulence of $\mathcal{M}_\text{turb} \sim 9$ and {$\beta \lesssim 0.1$}, the timescale for the dissipation of kinetic energy at shocks would be comparable to $t_\text{cross}$. \citet{Smith2000a,Smith2000b} calculated the amount of the energy dissipated by ``artificial viscosity'' at shocks, and suggested that this would be $\sim 1/2$ and $\sim 2/3$ of the total energy dissipation for decaying and driven supersonic turbulences, respectively. All these numerical works and observations hint that the shock dissipation would account for a fraction of the turbulent energy dissipation, although it has not been preciously defined.

\begin{deluxetable*}{l|ccccccc}
\tablecaption{List of MHD Turbulence Simulations \label{Table1}}
\tablehead{Model Name $^a$ & $N_\text{fa}/n_\text{g}^2$ $^b$ & $N_\text{sl}/n_\text{g}^2$ $^b$ & $\epsilon_\text{inj}$ $^c$ & $\epsilon_\text{fa}/\epsilon_\text{inj}$ $^d$& $\epsilon_\text{sl}/\epsilon_\text{inj}$ $^d$ & $\epsilon_\text{cas}/\epsilon_\text{inj}$ $^e$ & {$\beta_\text{sat}$ $^f$}}
\startdata
512M0.5-b0.1  &	0.0018	&	0.103	&	1.20	&	0.0122	&	0.0004	&	0.807	&	{0.099} 	\\	
512M0.5-b1 	  &	0.749	&	0.154	&	1.32	&	0.538	&	0.0691	&	0.715	&	{0.932} 	\\	
512M0.5-b10   &	3.38	    &	1.22  	&	1.68	&	0.486	&	0.0192	&	0.567	&	{5.580} 	\\	
\hline
512M1-b0.1 	  &	0.0520	&	0.303	&	1.30	&	0.0768	&	0.0144	&	0.728	&	{0.097} 	\\	
512M1-b1   	  &	3.34 	&	2.55	    &	1.70	&	0.472	&	0.224	&	0.571	&	{0.762} 	\\	
512M1-b10 	  &	7.77	    &	4.52	    &	2.17	&	0.390	&	0.0742	&	0.458	&	{2.420} 	\\	
\hline
512M2-b0.1 	  &	0.480	&	1.06	    &	1.50	&	0.135	&	0.0636	&	0.622	&	{0.091} 	\\	
512M2-b1   	  &	6.16	     &	4.09	    &	1.88	&	0.285	&	0.0867	&	0.467	&	{0.482} 	\\	
512M2-b10 	  &	11.8  	&	4.42 	&	1.94	&	0.328	&	0.0400	&	0.469	&	{1.150} 	\\	
\hline
512M4-b0.1 	  &	1.58	    &	1.57  	&	1.58	&	0.111	&	0.0215	&	0.555	&	{0.073} 	\\	
512M4-b1  	  &	7.32	    &	2.54	    &	1.81	&	0.192	&	0.0159	&	0.478	&	{0.230} 	\\	
512M4-b10 	  &	13.7  	&	2.59  	&	1.55	&	0.275	&	0.0107	&	0.556	&	{0.537} 	\\	
\hline
512M7-b0.1 	  &	2.92	    &	0.968	&	1.73	&	0.104	&	0.0034	&	0.494	&	{0.048} 	\\	
512M7-b1  	  &	9.15	    &	1.34 	&	1.72	&	0.193	&	0.0028	&	0.536	&	{0.123} 	\\	
512M7-b10 	  &	16.3	    &	1.48 	&	1.50	&	0.244	&	0.0020	&	0.618	&	{0.332} 	\\	
\hline
1024M0.5-b0.1 &	0.0011	&	0.0763	&	0.960&	0.0069	&	0.0002	&	0.975	&	{0.099} 	\\	
1024M0.5-b1   &	0.614	&	0.149	&	1.14	&	0.476	&	0.0951	&	0.803	&	{0.931} 	\\	
1024M0.5-b10  &	2.71 	&	0.88	    &	1.48	&	0.399	&	0.0262	&	0.600    &	{5.550} 	\\	
\hline
1024M1-b0.1   &	0.0277	&	0.242	&	1.10	&	0.0362	&	0.0236	&	0.839	&	{0.097} 	\\	
1024M1-b1 	  &	2.67	    &	2.66	    &	1.50	&	0.357	&	0.243	&	0.606	&	{0.771} 	\\	
1024M1-b10 	  &	6.68	    &	4.98	    &	2.09	&	0.292	&	0.0943	&	0.478	&	{2.330} 	\\	
\hline
1024M7-b0.1   &	2.55   	&	2.05   	&	1.79	&	0.0647	&	0.0059	&	0.484	&	{0.045} 	\\	
1024M7-b1 	  &	8.08  	&	2.55	    &	1.77	&	0.138	&	0.0048	&	0.510	&	{0.099} 	\\	
1024M7-b10 	  &	18.3 	&	2.74	    &	1.53	&	0.221	&	0.0034	&	0.585	&	{0.252} 	\\	
\enddata
\tablenotetext{a}{The starting number is $n_\text{g}$, the number after M is $\mathcal{M}_\text{turb}$, and the number after b is the initial plasma beta, $\beta_0$; $t_\text{end} = 6$ and $4.8 \times t_\text{cross}$ for $512^3$ and $1024^3$ simulations, respectively.}
\tablenotetext{b}{The number of grid zones identified as fast and slow shocks, normalized to $n_\text{g}^2$.}
\tablenotetext{c}{The energy injection rate in units of $\rho_0L_0^2c_\text{s}^3\mathcal{M}_\text{turb}^3$.}
\tablenotetext{d}{The fraction of the energy dissipation at fast and slow shocks.}
\tablenotetext{e}{The fraction of the energy dissipation through turbulent cascade, estimated with the flow crossing time. See Section \ref{sec3.3} for the definition.}
{\tablenotetext{f}{The plasma beta calculated with $\left< B^2 \right>_\text{sat}$ of the saturated stage.}}
\end{deluxetable*}

In this paper, we study shocks in isothermal, compressible, driven MHD turbulence, focusing on their Mach number probability distribution and spatial frequency, as well as the energy dissipation at shocks. Wide ranges of $\mathcal{M}_\text{turb}$ and $\beta$ are considered, intending to cover turbulences in the ISM and ICM. Specifically, we present an algorithm to identify shocked grid zones, separate them into fast and slow shocks, and calculate the Mach numbers. In particular, for the first time, we explicitly define the energy dissipation at shocks and introduce formulae to calculate it. We then apply those to a homogeneous set of simulation data with up to $1024^3$ grid zones.

The paper is organized as follows. Section \ref{sec2} describes the details of numerics, including the parameters for turbulence simulations. Section \ref{sec3} presents the algorithm for shock identification and the formulae for the calculations of shock Mach number and energy dissipation at shocks. The main results, i.e., the shock statistics and the energy dissipation at shocks, are given in Section \ref{sec4}. Summary and discussion follow in Section \ref{sec5}.

\section{Numerics}\label{sec2}

\begin{figure*}
\begin{center}
\vskip 0.2cm
\hskip -1.0cm
\includegraphics[width=1.05\textwidth]{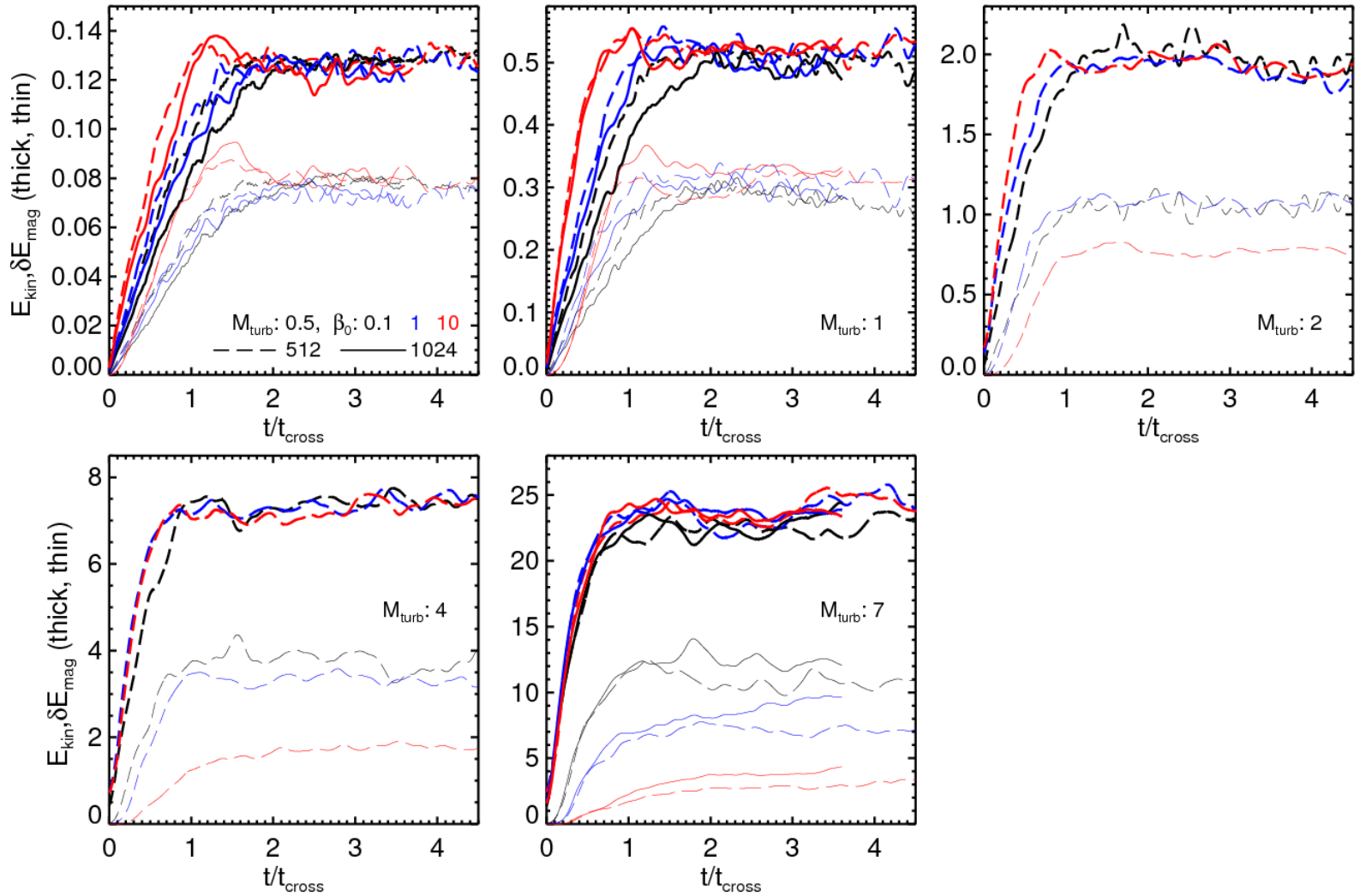}
\vskip -0.1cm
\caption{Time evolution of the kinetic energy, $E_\text{kin}$ (thick lines), and the magnetic energy increase, $\delta E_\text{mag} = (\bm{B} - \bm{B}_0)^2 / 2$ (thin lines), for the simulations listed in Table \ref{Table1}.} \label{fig1}
\end{center} 
\end{figure*}

Turbulence simulations solved the following set of equations for isothermal, compressible MHDs,
\begin{eqnarray}
\frac{\partial\rho}{\partial{t}}&&+\bm{\nabla}\cdot\left(\rho\bm{v}\right)=0, \label{eq1}\\
\frac{\partial\bm{v}}{\partial{t}}+\bm{v}\cdot\bm{\nabla}\bm{v}+&&\frac{1}{\rho}\bm{\nabla}{P}-\frac{1}{\rho}\left(\bm{\nabla}\times\bm{B}\right)\times\bm{B}=0, \label{eq2}\\
\frac{\partial\bm{B}}{\partial{t}}&&-\bm{\nabla}\times\left(\bm{v}\times\bm{B}\right)=0, \label{eq3}
\end{eqnarray}
where the unit of $\bm{B}$ was chosen so that $4\pi$ does not appear in Equation (\ref{eq2}). The gas pressure is given as $P \equiv \rho c_\text{s}^2$ with a constant sound speed $c_\text{s}$ (the isothermal condition). A multi-dimensional code described in \citet{Kim1999} was employed for simulations. It is based on the explicit, finite-difference ``Total Variation Diminishing" (TVD) scheme, which is a second-order accurate upwind, and enforces $\bm{\nabla}\cdot\bm{B}=0$ with a constrained transport scheme \citep{Ryu1998}. The code does not explicitly model viscous and resistive dissipations.

Simulations were performed in a three-dimensional (3D) periodic box of size $L_0=1$ using $n^3_\text{g}=512^3$ and $1024^3$ grid zones. Initially, the medium was uniform with $\rho_0=1$, $P_{g,0}=1$ (so $c_\text{s}=1$), and $\bm{B}_0=(B_0,0,0)$, and at rest with $\bm{v} = 0$. Turbulence was driven with traditional, ``solenoidal'' forcing ($\bm{\nabla}\cdot\delta\bm{v}=0$) \citep[see, e.g.,][]{Stone1998,Mac1999}. {We note that the properties of turbulence depends on the nature of forcing. \citet{Porter2015}, for instance, demonstrated differences in ICM turbulence when solenoidal or compressive ($\bm{\nabla}\times\delta\bm{v}=0$) forcings were applied. And \citet{federrath2010} argued that for modeling of turbulence in MCs, a mixture of solenoidal and compressive drivings would be natural, based on degrees of freedom arguments. As the first trial to quantify the statistics and dissipation of shocks, we here consider turbulence driven by solenoidal forcing.}

Velocity perturbations, $\delta\bm{v}$, were drawn from a Gaussian random field determined with power spectrum, $|\delta v_k|^2 \propto k^6\text{exp}(-8k/k_\text{exp})$, where $k_\text{exp}=2\ k_0$ $(k_0=2\pi/L_0)$, and added to $\delta \bm{v}$ at {each interval of} $\Delta t=0.001L_0/c_\text{s}$. The amplitude of the perturbations was tuned in such a way that the desired $v_\text{turb} \equiv v_\text{rms} = \left< v^2 \right>^{1/2}$ is achieved. Following previous works \citep[e.g.,][]{Stone1998}, the injection scale is set as that of $k_\text{exp}$, i.e., $L_\text{inj} = L_0/2$, although the forcing has a peak around $\sim 1.5\ k_0$.

The problem is then defined by two parameters, the turbulent Mach number, $\mathcal{M}_\text{turb} \equiv v_\text{rms}/c_\text{s}$, and the initial plasma beta, $\beta_0 \equiv P_{g,0}/P_{B,0} = \rho_0 c_s^2 / (B_0^2/2)$. To cover turbulences in different astrophysical environments, as mentioned in the Introduction, the cases of $\mathcal{M}_\text{turb} = 0.5$ (subsonic turbulence), 1 (transonic turbulence), and 2, 4, 7 (supersonic turbulence), and $\beta_0=$ 0.1 (strong field), 1 (moderate field), and 10 (weak field) were considered. Table \ref{Table1} lists the simulations performed for this paper, along with their model names. High-resolution simulations using $1024^3$ grid zones were made only for $\mathcal{M}_\text{turb}=0.5,\;1,\;7$, while those of $512^3$ grid zones were done for all the cases. Figure \ref{fig1} shows the kinetic and magnetic energy evolutions in those simulations; simulations ran up to $t = 6$ and $4.8 \times t_\text{cross}$ for $512^3$ and $1024^3$ simulations, respectively, where $t_\text{cross} = L_\text{inj}/ v_\text{rms}$. The kinetic energy glows quickly within $\sim 1\ t_\text{cross}$, as was shown in previous works \citep[see, e.g.,][]{federrath2013,Porter2015}. The growth of the magnetic energy follows and reaches saturation by $\sim 2\ t_\text{cross}$. {The plasma beta at the saturated stage, $\beta_\text{sat}$, is listed in Table \ref{Table1}.} The results presented in Section \ref{sec4} were drawn from the data after the saturation, specifically, during $t/t_\text{cross} = 2 - 6$ and $2 - 4.8$ for $512^3$ and $1024^3$ simulations, respectively.

\section{Methods}\label{sec3}

\subsection{Identification of Shocks}\label{sec3.1}

In simulation data, shocks (actually grid zones that are parts of shock surfaces) were identified with the following algorithm. Along each coordinate direction, grid zones were tagged as ``shocked'', if $\bm{\nabla}\cdot\bm v<0$, i.e., the local flow is converging, and $\max(\rho_{i+1}/\rho_{i-1},\rho_{i-1}/\rho_{i+1}) \geq 1.03^2$ around the zone $i$. The second condition, corresponding to the density jump of isothermal ``hydrodynamic'' shocks with sonic Mach numbers $\geq 1.03$ (see Section \ref{sec3.2}), excludes weak shocks; it was imposed to avoid confusions between very weak shocks and waves. In simulations, shocks are spread and captured over a few to several grid zones. The ``shock center'' was identified as the zone with minimum $\bm{\nabla}\cdot\bm v$ among attached shocked zones. The preshock and postshock zones were defined around the center zone, assuming the spread of shocks is typically over 2 to 4 zones. {Specifically, for a shock center $i$, $i+2$ and $i-2$ were chosen as either preshock or postshock zones, if $i+1$ and $i-1$ were shocked zones; otherwise, $i+1$ or $i-1$ were chosen.}

Once shock centers were identified, the sonic Mach number, $M_\text{s}$, was calculated using the flow quantities at the preshock and postshock zones with the formulae presented in the next subsection (Equations \ref{eq9}). If a zone was identified as shock centers along more than one-directions, the maximum value of $M_\text{s}$ was taken as its Mach number; that is, $M_\text{s}=\max(M_{\text{s},x},M_{\text{s},y},M_{\text{s},z})$ if identified as shock centers in all coordinate directions.

The above algorithm is similar to the one used before with the data for cosmological, large-scale structure formation simulations \citep[e.g.,][]{Ryu2003,Ha2018}, but different from the one applied to MHD turbulence data in \citet{Porter2015}. {The current algorithm, which is based on the dimension-per-dimension identification of shocks, is simpler than the one in \citet{Porter2015}, which uses boxes of $5 \times 5\times 5$ zones and identify shocks in 3D by finding the principle direction of density variation.} Yet both algorithms produce almost identical results, as already noted in \citet{Porter2015}.

\subsection{Calculation of Shock Mach Number}\label{sec3.2}

\begin{deluxetable*}{l|ccc|ccc}
{\tablecaption{Properties of identified shocks\label{Table2}}
\tablehead{\multirow{2}{*}{Model Name $^a$}&\multicolumn{3}{c|}{fast shocks}&\multicolumn{3}{c}{slow shocks}\\
&$\left< M_\text{s} \right>$ $^b$ &$\left< \rho_1 \right>$ $^c$ &$\left< B_1 \right>$ $^d$ &$\left< M_s \right>$ $^b$ &$\left< \rho_1 \right>$ $^c$ &$\left< B_1 \right>$ $^d$}
\startdata
512M0.5-b0.1  &	4.630 	&	1.310 	&	1.031 	&	1.120 	&	0.842 	&	1.013 	\\	
512M0.5-b1 	  &	1.850 	&	1.100 	&	0.948 	&	0.806 	&	0.922 	&	1.068 	\\	
512M0.5-b10   &	1.250 	&	1.000 	&	1.071 	&	0.660 	&	0.901 	&	1.919 	\\	\hline
512M1-b0.1 	  &	4.330 	&	1.700 	&	0.991 	&	1.160 	&	0.870 	&	1.013 	\\	
512M1-b1 	  &	1.900 	&	1.170 	&	0.884 	&	0.848 	&	0.936 	&	1.181 	\\	
512M1-b10 	  &	1.460 	&	0.968 	&	1.310 	&	0.788 	&	0.881 	&	2.594 	\\	\hline
512M2-b0.1 	  &	3.870 	&	2.200 	&	0.870 	&	1.100 	&	1.150 	&	1.029 	\\	
512M2-b1 	  &	2.240 	&	1.200 	&	0.898 	&	0.938 	&	1.000 	&	1.471 	\\	
512M2-b10 	  &	1.900 	&	0.940 	&	1.693 	&	0.899 	&	0.976 	&	3.466 	\\	\hline
512M4-b0.1 	  &	4.210 	&	2.420 	&	0.823 	&	1.130 	&	1.490 	&	1.116 	\\	
512M4-b1 	  &	3.250 	&	1.300 	&	1.188 	&	1.050 	&	1.250 	&	1.994 	\\	
512M4-b10 	  &	2.750 	&	0.956 	&	2.258 	&	1.010 	&	1.190 	&	4.584 	\\	\hline
512M7-b0.1 	  &	5.880 	&	2.150 	&	0.910 	&	1.170 	&	1.920 	&	1.299 	\\	
512M7-b1 	  &	5.000 	&	1.280 	&	1.570 	&	1.130 	&	1.560 	&	2.475 	\\	
512M7-b10 	  &	3.930 	&	0.972 	&	2.773 	&	1.080 	&	1.430 	&	5.098 	\\	\hline
1024M0.5-b0.1 &	4.740 	&	1.260 	&	1.038 	&	1.120 	&	0.818 	&	1.008 	\\	
1024M0.5-b1   &	1.870 	&	1.120 	&	0.976 	&	0.668 	&	0.898 	&	1.082 	\\	
1024M0.5-b10  &	1.250 	&	1.030 	&	1.091 	&	0.585 	&	0.912 	&	1.876 	\\	\hline
1024M1-b0.1   &	4.290 	&	1.730 	&	1.011 	&	1.120 	&	0.878 	&	1.017 	\\	
1024M1-b1 	  &	1.870 	&	1.220 	&	0.891 	&	0.757 	&	0.947 	&	1.181 	\\	
1024M1-b10 	  &	1.440 	&	1.020 	&	1.348 	&	0.726 	&	0.901 	&	2.616 	\\	\hline
1024M7-b0.1   &	5.040 	&	3.190 	&	0.939 	&	1.070 	&	2.110 	&	1.389 	\\	
1024M7-b1 	  &	4.660 	&	1.760 	&	1.824 	&	1.050 	&	1.650 	&	2.864 	\\	
1024M7-b10 	  &	3.760 	&	1.160 	&	3.287 	&	1.010 	&	1.480 	&	5.836 	\\
\enddata
\tablenotetext{a}{See Table \ref{Table1} for the convention of model name and the model parameters.}
\tablenotetext{b}{The average sonic Mach number of identified fast and show shocks.}
\tablenotetext{c}{The average preshock density of identified fast and show shocks in units of $\rho_0$.}
\tablenotetext{d}{The average preshock magnetic field strength of identified fast and show shocks in units of $B_0$.} }
\end{deluxetable*}

\begin{figure*}
\begin{center}
\vskip 0.2cm
\hskip -0.2cm
\includegraphics[width=1\textwidth]{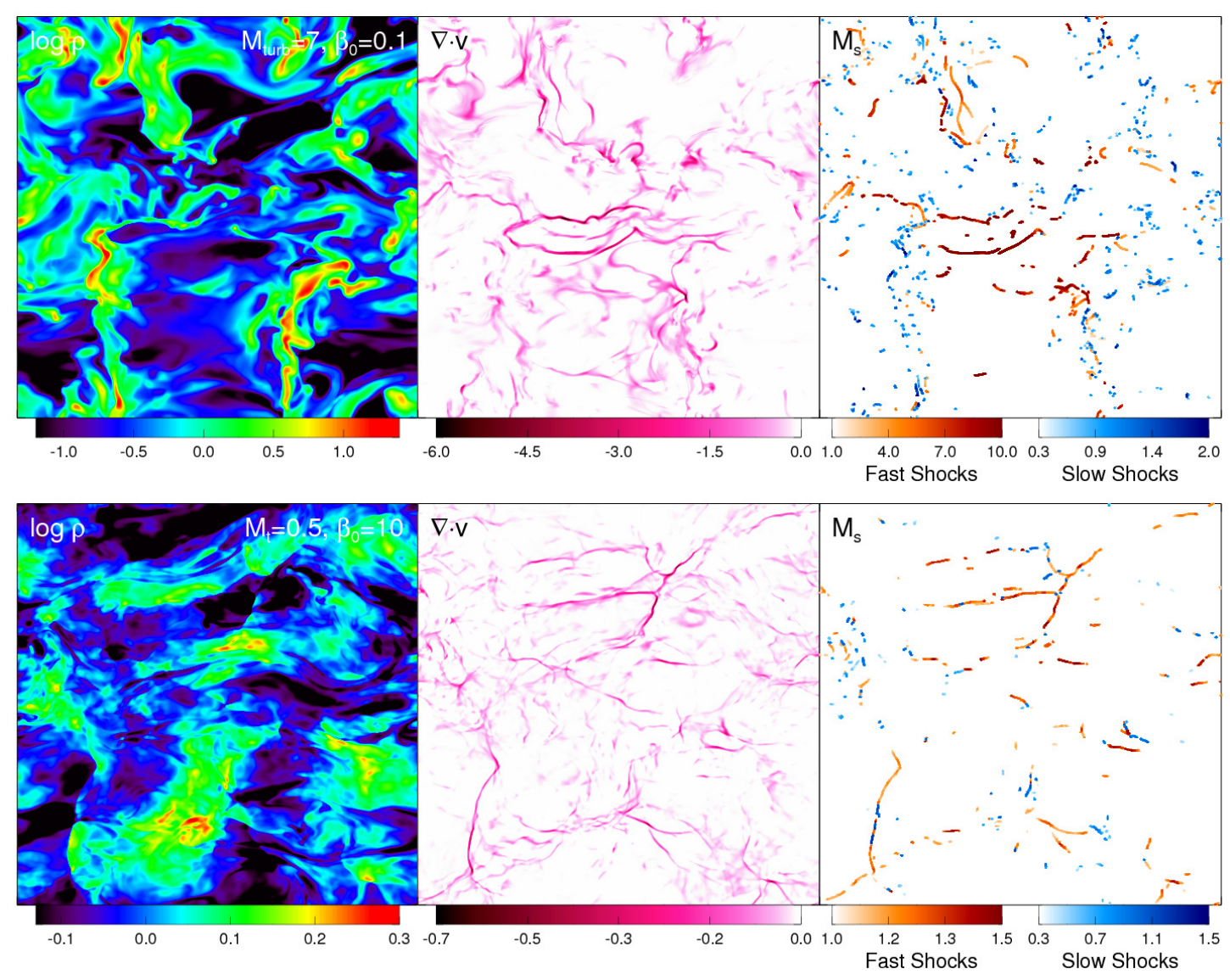}
\vskip -0.2cm
\caption{2D slice images of log $\rho$, $\bm{\nabla}\cdot\bm{v}$, and $M_\text{s}$ for fast and slow shocks at $t_\text{end}$ in simulations using $1024^3$ grid zones of $\mathcal{M}_\text{turb}=7$ and $\beta_0=0.1$ (1024M7-b0.1, top panels) and $\mathcal{M}_\text{turb}=0.5$ and $\beta_0=10$ (1024M0.5-b10, bottom panels). The mean magnetic field $\bm{B}_0$ is oriented along the horizontal axis.} \label{fig2}
\end{center} 
\end{figure*}

\begin{figure*}
\begin{center}
\hskip -0.2cm
\includegraphics[width=1\textwidth]{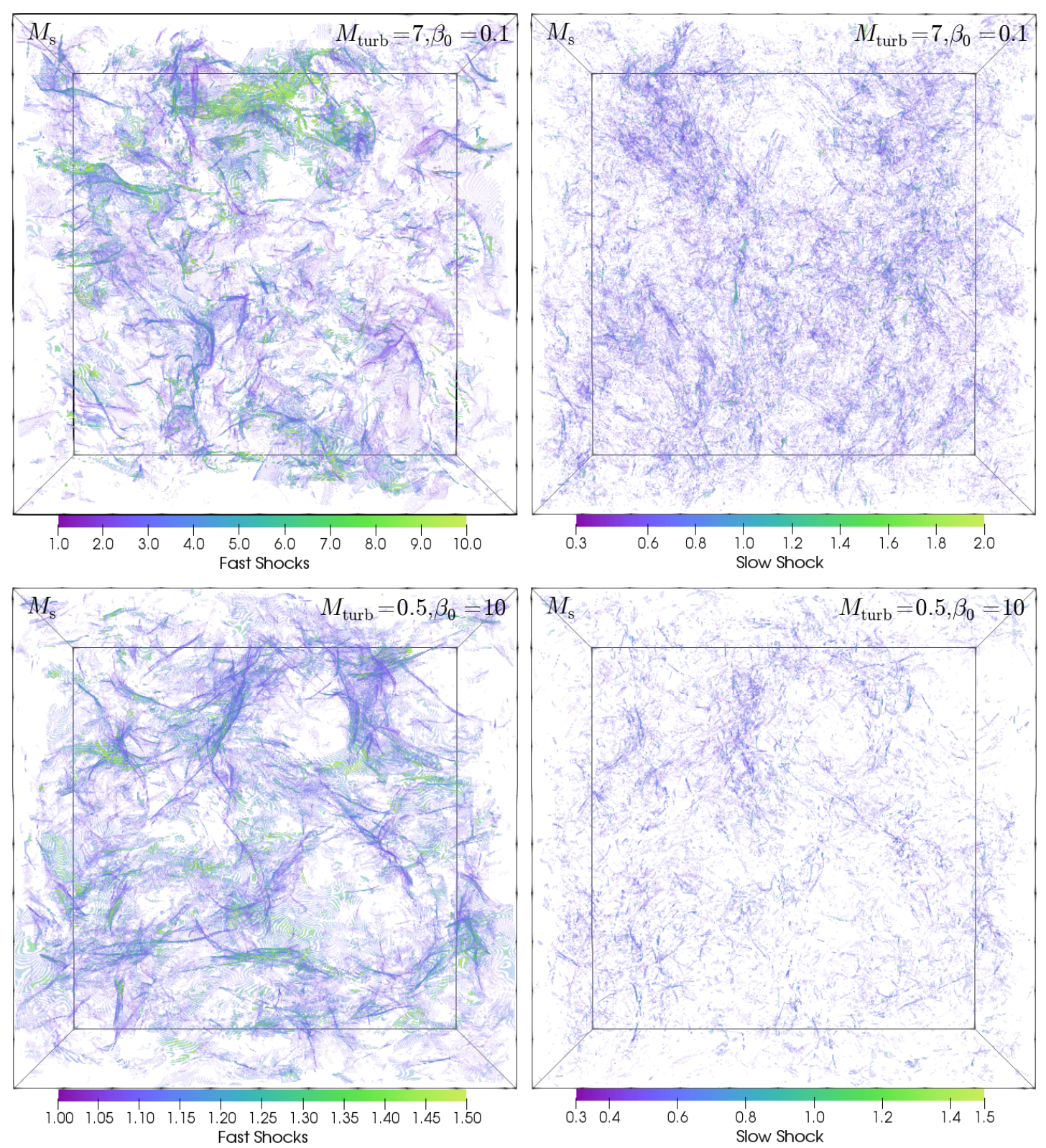}
\vskip -0.2cm
\caption{3D distributions of fast shocks (left panels) and slow shocks (right panels) at $t_\text{end}$ in simulations using $1024^3$ grid zones of $\mathcal{M}_\text{turb}=7$ and $\beta_0=0.1$ (1024M7-b0.1, top panels) and $\mathcal{M}_\text{turb}=0.5$ and $\beta_0=10$ (1024M0.5-b10, bottom panels). The mean magnetic field $\bm{B}_0$ is oriented along the horizontal axis.} \label{fig3} 
\end{center}
\end{figure*}

Formulae to calculate $M_\text{s}$ can be derived from the shock jump conditions. From the conservative form of Equations (\ref{eq1}) - (\ref{eq3}), the jump conditions for isothermal flows in the {\it shock-rest frame} are written as
\begin{eqnarray} 
\rho_1v_{\parallel1}&=&\rho_2v_{\parallel2}, \label{eq4}\\
\rho_1v_{\parallel1}^2+c_\text{s}^2\rho_1+\frac{1}{2}B_{\perp1}^2&=&\rho_2v_{\parallel2}^2+c_\text{s}^2\rho_2+\frac{1}{2}B_{\perp2}^2, \label{eq5}\\
\rho_1v_{\parallel1}v_{\perp1}-B_{\parallel}B_{\perp1}&=&\rho_2v_{\parallel2}v_{\perp2}-B_{\parallel}B_{\perp2}, \label{eq6}\\
v_{\parallel1}B_{\perp1}-v_{\perp1}B_{\parallel}&=&v_{\parallel2}B_{\perp2}-v_{\perp2}B_{\parallel}, \label{eq7}
\end{eqnarray}
where the subscripts 1 and 2 indicate the preshock and postshock quantities, respectively, and $\parallel$ and $\perp$ denote the components parallel and perpendicular to the shock normal, respectively. The parallel field is continuous across the shock, so $B_{\parallel1} = B_{\parallel2} = B_{\parallel}$.

The manipulation of the jump conditions becomes substantially simplified in the so-called {\it preferred frame}, in which $\bm{v}$ and $\bm{B}$ are parallel to one another on both sides of the shock, i.e., ${B_{\perp1}}/{B_{\parallel}}={v_{\perp1}}/{v_{\parallel1}}$ and ${B_{\perp2}}/{B_{\parallel}}={v_{\perp2}}/{v_{\parallel2}}$ \citep[see, e.g.,][]{shu1992}. In the frame, Equation (\ref{eq7}) is automatically satisfied. By combining Equations (\ref{eq4}) - (\ref{eq6}), we can get, for instance,
\begin{eqnarray}
&&M_\text{s}^6 - \left[\left(1+\frac{2 v_{\text{A}\parallel}^2}{c_\text{s}^2}+\frac{v_{\text{A}\perp1}^2}{2 c_\text{s}^2}\right)\chi+\frac{v_{\text{A}\perp1}^2}{2 c_\text{s}^2}\chi^2\right]M_\text{s}^4 \nonumber \\
&&+\frac{v_{\text{A}\parallel}^2}{c_\text{s}^2}\left(2+\frac{v_{\text{A}\parallel}^2}{c_\text{s}^2}+\frac{v_{\text{A}\perp1}^2}{c_\text{s}^2}\right)\chi^2 M_\text{s}^2 - \frac{v_{\text{A}\parallel}^4}{c_\text{s}^4}\chi^3=0, \label{eq8}
\end{eqnarray}
where $M_\text{s} = v_{\parallel1}/c_\text{s}$, $\chi={\rho_2}/{\rho_1}={v_{\parallel1}}/{v_{\parallel2}}$, $v_{\text{A}\perp1}={B_{\perp1}}/{\sqrt{\rho_1}}$, and $v_{\text{A}\parallel}={B_{\parallel}}/{\sqrt{\rho_1}}$. The equation is written in such a way that the shock sonic Mach number, $M_\text{s}$, is calculated with the density compression and the preshock magnetic field. Note that the flow velocity in the shock-rest, preferred frame is different from that in the computational frame, so $\bm{v}$ from simulation data can not be used to compute $M_\text{s}$ with formulae derived in the shock-rest, preferred frame. The magnetic field, on the other hand, does not change by coordinate transformations, as long as $v \ll c$ (the speed of light), which is the case for the turbulences in the ISM and ICM.

Equation (\ref{eq8}) is, however, a cubic equation for $M_\text{s}^2$, and solving it is rather complicated. Instead, a simpler formula, for instance,
\begin{equation}
M_\text{s}^2 = \chi+\frac{\chi}{\chi-1}\frac{B_2^2-B_1^2}{2 c_\text{s}^2 \rho_1}, \label{eq9}
\end{equation}
can be derived from Equations (\ref{eq4}) and (\ref{eq5}). This equation requires the preshock and postshock densities and magnetic fields, but those quantities can be extracted from simulation data. We employed Equation (\ref{eq9}) for the calculation of $M_\text{s}$. For isothermal hydrodynamic shocks with $\bm{B}=0$, both Equations (\ref{eq8}) and (\ref{eq9}) reduce to $M_\text{s}^2 = \chi$, as expected.

In MHDs, shocks appear as fast and slow modes; the strength of the perpendicular magnetic field increases across the fast shock, i.e., $B_{\perp2} > B_{\perp1}$, while it decreases across the slow shock, i.e., $B_{\perp12} < B_{\perp1}$ {\citep[see, e.g.,][]{shu1992}. In addition, the MHD equations accommodate shocks of the third mode, called ''intermediate shocks'', which are, however, known to be non-evolutionary or unphysical, particularly in ideal MHDs \citep[see, e.g.,][]{landau84}. The intermediate mode is manifested as rotational discontinuity in simulation \citep[see, e.g.,][]{ryu1995}. Hence, we did not seek intermediate shocks.}

Identified shocks were classified into either fast or slow populations according to the criterion of $B_{\perp2} > B_{\perp1}$ or $B_{\perp2} < B_{\perp1}$, and the statistics of the two populations were calculated separately. The speeds of fast and slow waves were calculated as
\begin{eqnarray}
c_\text{fa,sl}^2&=&\frac{1}{2}\left(c_\text{s}^2+v_{\text{A}\parallel}^2+v_{\text{A}\perp}^2\right) \nonumber  \\  
&&\pm\frac{1}{2}\sqrt{\left(c_\text{s}^2+
v_{\text{A}\parallel}^2+v_{\text{A}\perp}^2 \right)^2-4v_{\text{A}\parallel}^2c_\text{s}^2}, \label{eq10}
\end{eqnarray}
with the $+$ and $-$ signs referring to the fast slow modes, respectively. {Here and below, we used the quantities along the direction of shock identification for those with $\parallel$, and the quantities perpendicular to the direction for those with $\perp$.} After the sonic Mach numbers of shocks, $M_\text{s}$, were obtained with Equation (\ref{eq9}), their fast and slow Mach numbers were calculated as $M_\text{fa}=M_\text{s}c_\text{s}/c_\text{fa,1}$ and $M_\text{sl}=M_\text{s}c_\text{s}/c_\text{sl,1}$, where $c_\text{fa,1}$ and $c_\text{sl,1}$ are the wave speeds in the preshock zones.

To avoid confusions from complex flow patterns and shock surface topologies, only shocks with $\max(\rho_{i+1}/\rho_{i-1},$ $\rho_{i-1}/\rho_{i+1}) \geq 1.03^2$ and $M_\text{fa} \geq 1.06$ (for fast shocks) and $M_\text{sl} \geq 1.06$ (for slow shocks) were used for the statistics presented in Section \ref{sec4}. Weak shocks with $M_\text{fa,sl} < 1.06$ are expected to dissipate little energy.

In general, the identification of fast shocks is relatively straightforward. Slow shocks, on the other hand, are harder to be reliably identified, mainly due to the following two reasons. First, {the surfaces of slow shocks could be subject to a corrugation instability} \citep[see, e.g.,][]{Stone1995}, and hence their surfaces are distorted and fragmented (see Section \ref{sec4.1}). Then, they could be easily confused with waves, especially when the shock normal is not aligned with coordinate axes. Second, some of slow shocks can have small shock speeds, $v_{\parallel1} \ll c_\text{s}$, when the preshock flows have small slow wave speeds, $c_\text{sl,1} \ll c_\text{s}$. In these cases, the distinction between shocks and fluctuations is often not very clear. So for slow shocks, an additional constraint, $c_\text{sl,1}/c_\text{s} \geq 0.3$, was imposed. In the cases of low $\mathcal{M}_\text{turb}$ and small $\beta_0$, slow shocks with $c_\text{sl,1}/c_\text{s} \lesssim 0.3$ mostly propagate perpendicular to the background magnetic field and have small $M_\text{s}$ (see Section \ref{sec4.2}). In the cases of high $\mathcal{M}_\text{turb}$, however, the constraint might exclude some shocks with substantial $M_\text{s}$. On the other hands, some of waves or fluctuations may have been counted as slow shocks. Hence, the population of slow shocks, presented in the next section, might have been estimated less accurately than that of fast shocks.

{Table \ref{Table2} lists some of the average quantities of identified fast and slow shocks. See the next section for discussions.}

In Appendix \ref{secA1}, the dependence of the statistics on the shock identification parameters, i.e., the constraints on $\max(\rho_{i+1}/\rho_{i-1},\rho_{i-1}/\rho_{i+1})$, $M_\text{fa,sl}$, and $c_\text{sl,1}$, is presented.

\subsection{Calculation of Energy Dissipation}\label{sec3.3}

As noted in the Introduction, some of the turbulent energy directly dissipates at shocks, while the rest cascades down and dissipates into heat. For adiabatic flows, the conservation equation for the total energy, composed of the kinetic, magnetic, and thermal energies, follows actions in the energy. In the case of isothermal flows, \citet{Mouschovias1974} showed that the ``heat energy'' or the ``effective internal energy'', $P \ln P$, which obeys
\begin{equation}
\frac{\partial P \ln P}{\partial{t}} + \bm{\nabla} \cdot \left(P \ln P \bm{v} \right) + P \bm{\nabla} \cdot \bm{v} = 0, \label{eq11}
\end{equation}
can be introduced, and the equation for the ``effective total energy'' can be written down as
\begin{eqnarray} 
&&\frac{\partial}{\partial{t}}\left(\frac{1}{2}\rho v^2+P\ln P+\frac{1}{2}B^2\right) + \nonumber \\
&&\bm{\nabla}\cdot\left[\left(\frac{1}{2}\rho v^2+P\ln P+P\right)\bm{v} + \left(\bm{B}\times\bm{v}\right)\times\bm{B}\right] \nonumber \\
&& = 0.  \label{eq12}
\end{eqnarray} 

The above equation holds in smooth parts of flows, but does not at shocks; it is because the isothermal equation of state assumes an instantaneous loss of the thermal energy converted from the kinetic and magnetic energies at shocks. The jump of the energy flux in the shock-rest frame,
\begin{eqnarray}
&&\left[\left(\frac{1}{2}\rho v^2+P\ln P+P+B^2\right)v_{\parallel}-B_{\parallel}\left(B_{\parallel}v_{\parallel}+B_{\perp}v_{\perp}\right)\right]^1_2 \nonumber\\
&&\equiv Q, \label{eq13}
\end{eqnarray}
hence, should estimate the energy lost at isothermal shocks. Here, $\left[f\right]^1_2=f_1-f_2$ denotes the jump between the preshock and postshock quantities. So we define $Q$ as the energy dissipation rate per unit area at shock surfaces.

In the preferred frame, $Q$ is simplified to
\begin{equation}
Q=\left(\frac{1}{2}\rho_1v_1^2+P_1 \ln P_1 \right)v_{\parallel1} -\left(\frac{1}{2}\rho_2v_2^2+P_2 \ln P_2 \right)v_{\parallel2}, \label{eq14}
\end{equation}
and further written as
\begin{eqnarray}
&&\frac{Q}{\rho_1 M_\text{s} c_\text{s}^3} = \nonumber \\
&&\frac{1}{2}M_\text{s}^2\left[1-\frac{1}{\chi^2}+\frac{v_{\text{A}\perp1}^2\left(\chi-1\right)\left\{v_{\text{A}\parallel}^2\left(\chi+1\right)-2M_\text{s}^2c_\text{s}^2\right\}}{\left(v_{\text{A}\parallel}^2\chi - M_\text{s}^2 c_\text{s}^2\right)^2}\right] \nonumber \\
&&-\ln \chi. \label{eq15} 
\end{eqnarray}
Here, $Q$ is expressed in terms of the shock Mach number and the flow quantities independent of coordinate transformations, and hence can be calculated with simulation data.

\begin{figure*}
\begin{center}
\vskip 0.2cm
\hskip -0.6cm
\includegraphics[width=1.03\textwidth]{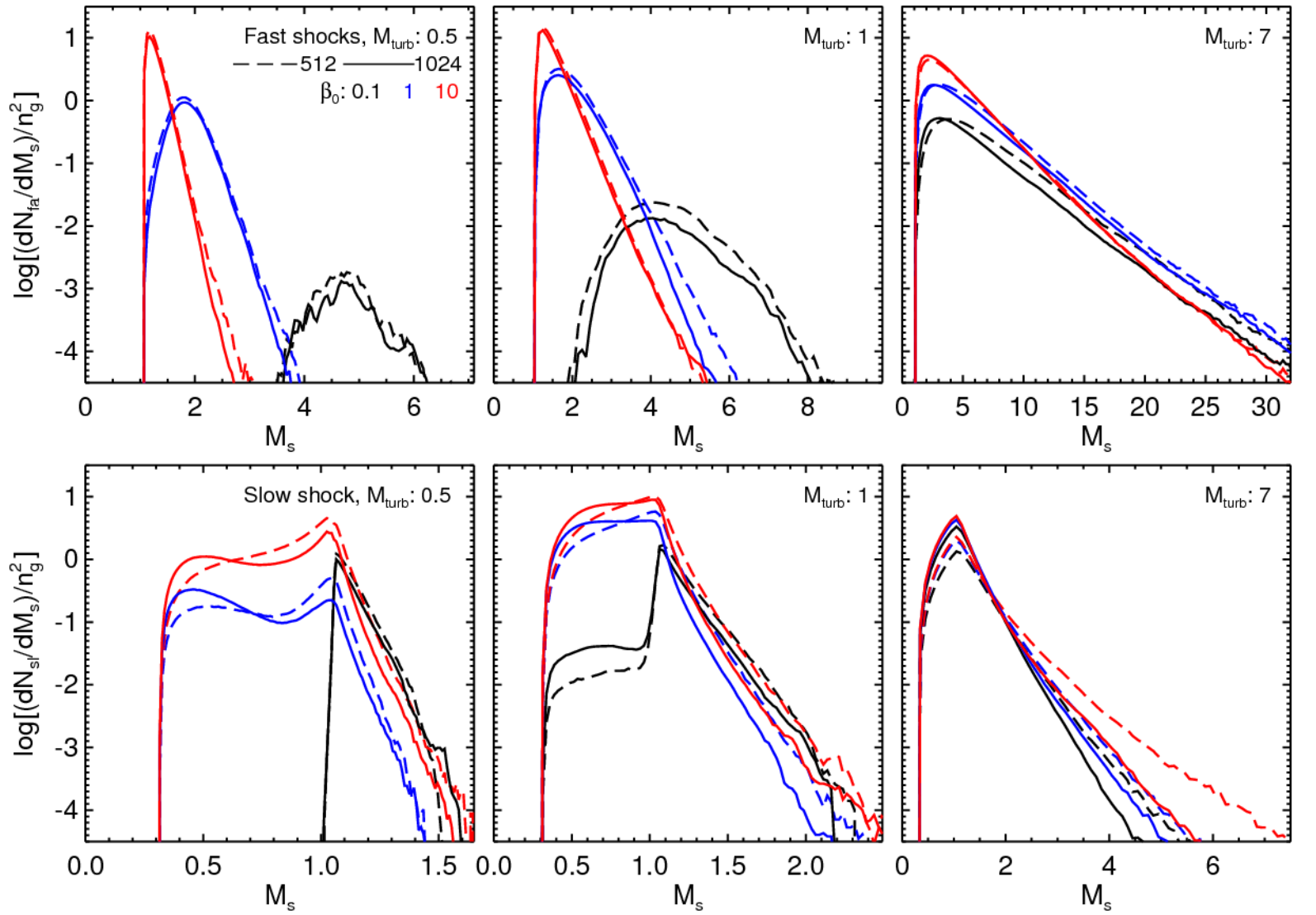}
\vskip -0.1cm
\caption{PDFs of sonic Mach number, $M_\text{s}$, for fast (upper panels) and slow (lower panels) shocks in turbulences with different $\beta_0$ and $\mathcal{M}_\text{turb}$. {The plasma beta at the saturated stage, $\beta_\text{sat}$, can be found in Table \ref{Table1}.} Note that while the vertical axes are drawn in the same scale, the horizontal axes cover different ranges of $M_\text{s}$.} \label{fig4}
\end{center}
\end{figure*}

\begin{figure}
\begin{center}
\vskip 0.2cm
\hskip -0.9cm
\includegraphics[width=0.52\textwidth]{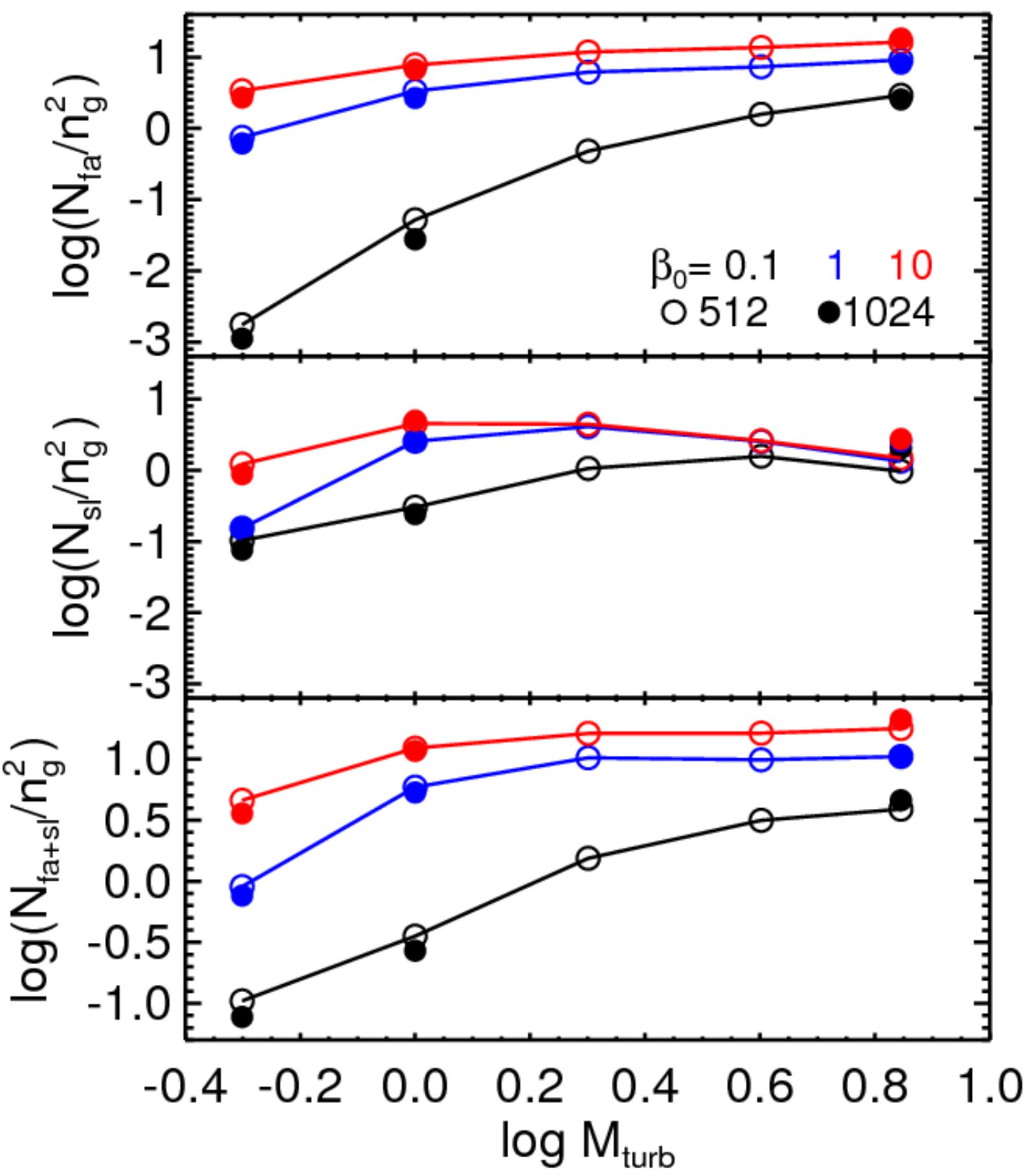}
\vskip -0.1cm
\caption{Number of the grid zones identified as fast shocks ($N_\text{fa}$, top) and slow shocks ($N_\text{sl}$, middle) and their sum ($N_\text{fa+sl}$, bottom), normalized to $n_\text{g}^2$, as a function of $\mathcal{M}_\text{turb}$ in turbulences with different $\beta_0$ and $\mathcal{M}_\text{turb}$. {The plasma beta at the saturated stage, $\beta_\text{sat}$, can be found in Table \ref{Table1}.}} \label{fig5}
\end{center}
\end{figure}

For hydrodynamic shocks, Equation (\ref{eq15}) reduces to
\begin{equation}
Q = \frac{1}{2} \rho_1 M_\text{s}^3 c_\text{s}^3 \left( \frac{M_\text{s}^4 - 1}{M_\text{s}^4} - \frac{4 \ln M_\text{s}}{M_\text{s}^2} \right). \label{eq16} 
\end{equation}
The term involving $\ln M_\text{s}$ inside the parenthesis (also the $-\ln \chi$ term in Equation (\ref{eq15})) originates from the heat energy term in Equation (\ref{eq12}), and so can be attributed to the assumption of isothermal flows. If flows are adiabatic and the gas cools instantaneously behind the shock, the energy lost at the shock is
\begin{equation}
Q=\frac{1}{2} \rho_1 v_{\parallel1}^3 - \frac{1}{2} \rho_2 v_{\parallel2}^3, \label{eq17}
\end{equation}
assuming the adiabatic index $\gamma = 1$ \citep[see, e.g.,][]{Ryu2003}; then, the first term inside the parenthesis in Equation (\ref{eq16}) is recovered.

With $Q$ in Equation (\ref{eq15}), we estimated the dissipation rate of turbulent energy at shocks, inside the whole computational box, as
\begin{equation}
\epsilon_\text{fa or sl}= \sum_{\text{fast or slow shocks}\ j} Q_j (\Delta x)^2, \label{eq18}
\end{equation}
separately for fast and slow shock populations. Here, $\Delta x=L_0/n_x$ is the size of grid zones, and the summation is over all the identified shock zones.

We also estimated the energy dissipation rate through turbulence cascade as
\begin{equation}
\epsilon_\text{cas} = \frac{1}{t_\text{cross}} \int \frac{1}{2}\rho v^2 dV, \label{eq19}
\end{equation}
where the integration is over the whole computational box \citep[see, e.g.,][]{Stone1998}. Note that if the total energy, i.e., the kinetic energy plus the magnetic energy increase, is used inside the integral, $\epsilon_\text{cas}$ would be $\sim 10 - 60\ \%$ larger (see Figure \ref{fig1}). The estimate would be ``independent'' of the scale and hence a fair value, only for incompressible Kolmogorov turbulence with spectral slope $5/3$. Hence, we here quote it only as a supplementary quantity.

The above energy dissipation rates were compared to the ``energy injection rate'', $\epsilon_\text{inj}$. Turbulence was driven by adding velocity perturbations, $\delta \bm{v}$, at intervals $\Delta t$, as described in Section \ref{sec2}. Hence,
\begin{equation} 
\epsilon_\text{inj} = \frac{1}{\Delta t}\int \frac{1}{2}\rho\left[\left(\bm v+\delta \bm v\right)^2-v^2\right] dV. \label{eq20}
\end{equation}
Below the energy dissipations, normalized to the energy injection, $\epsilon_\text{sh}/\epsilon_\text{inj}$ and $\epsilon_\text{cas}/\epsilon_\text{inj}$, are presented (also in Table \ref{Table1}).

\section{Results}\label{sec4}

\subsection{Spatial Distribution of Shocks}\label{sec4.1}

\begin{figure*}
\begin{center}
\vskip 0.2cm
\hskip -0.6cm
\includegraphics[width=1.03\textwidth]{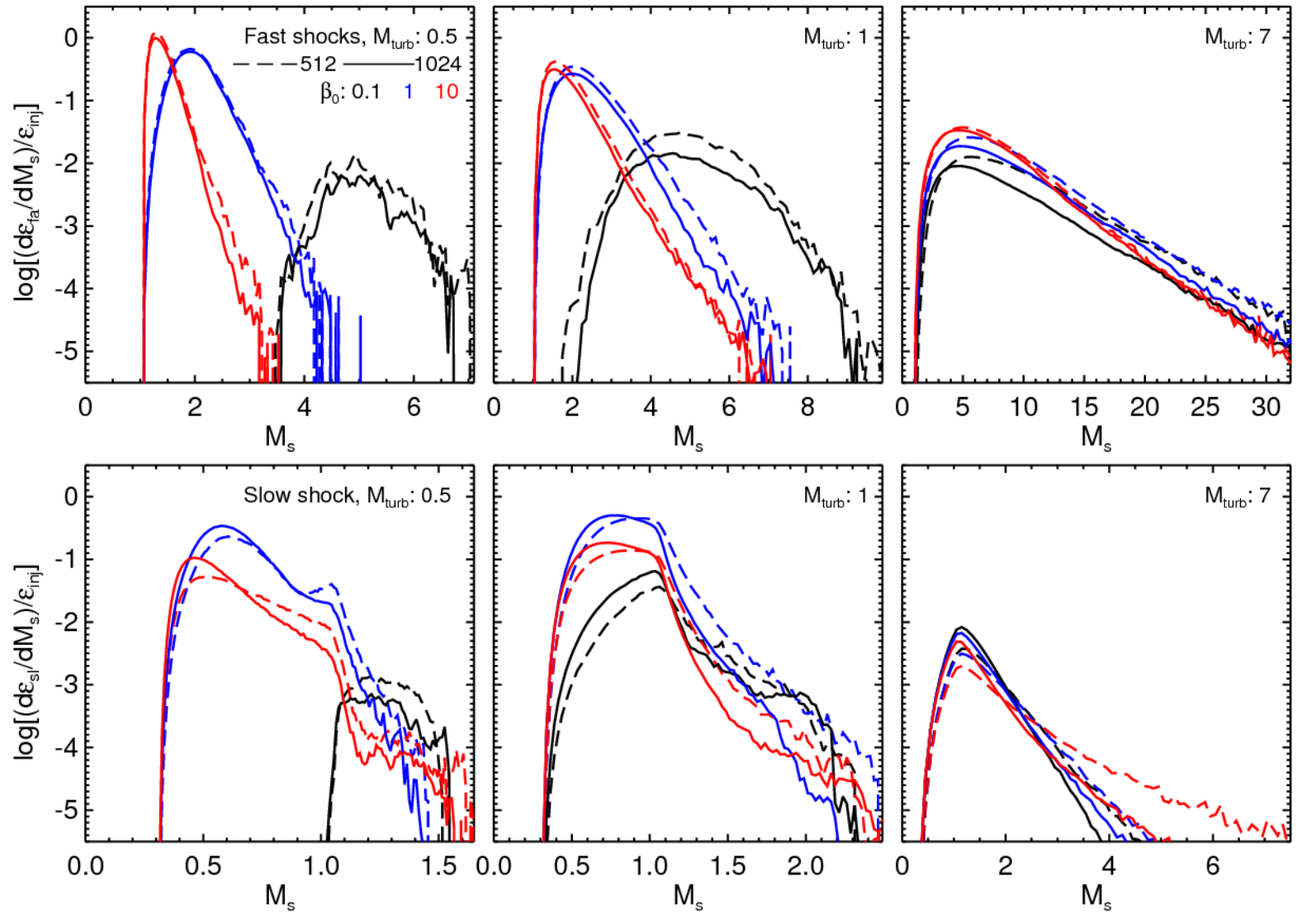}
\vskip -0.1cm
\caption{Energy dissipation rate distributions, normalized to $\epsilon_\text{inj}$, as a function of sonic Mach number, $M_\text{s}$, at fast (upper panels) and slow (lower panels) shocks in turbulences with different $\beta_0$ and $\mathcal{M}_\text{turb}$. {The plasma beta at the saturated stage, $\beta_\text{sat}$, can be found in Table \ref{Table1}.}. Note that while the vertical axes are drawn in the same scale, the horizontal axes cover different ranges of $M_\text{s}$.} \label{fig6}
\end{center}
\end{figure*}

\begin{figure}
\begin{center}
\vskip 0.2cm
\hskip -0.9cm
\includegraphics[width=0.52\textwidth]{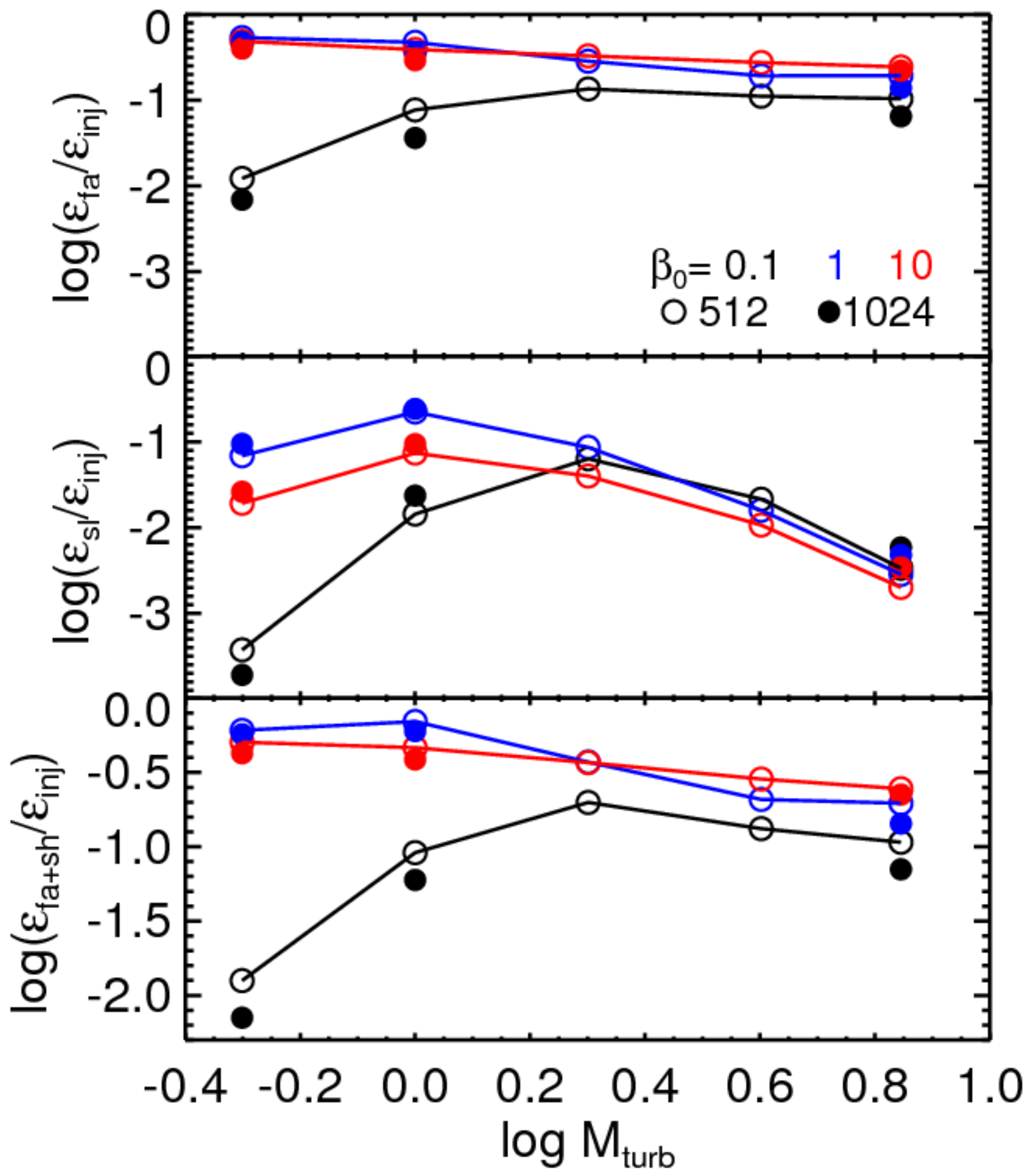}
\vskip -0.1cm
\caption{Fraction of the energy dissipation at fast shocks ($\epsilon_\text{fa}$, top) and at slow shocks ($\epsilon_\text{sl}$, middle) and their sum ($\epsilon_\text{fa+sl}$, bottom) as a function of $\mathcal{M}_\text{turb}$ in turbulences with different $\beta_0$ and $\mathcal{M}_\text{turb}$. {The plasma beta at the saturated stage, $\beta_\text{sat}$, can be found in Table \ref{Table1}.}} \label{fig7}
\end{center}
\end{figure}

Figure \ref{fig2} shows the spatial distributions of $M_\text{s}$ for fast and slow shocks, along with the distributions of the density and the flow convergence ($\bm{\nabla}\cdot\bm{v}$), in a two-dimensional (2D) slice, for 1024M7-b0.1 with $\mathcal{M}_\text{turb}=7$ and $\beta_0=0.1$, which intends to reproduce the turbulence in MCs, and for 1024M0.5-b10 with $\mathcal{M}_\text{turb}=0.5$ and $\beta_0=10$, which targets the turbulence in the ICM, at $t_\text{end}$. The density images display the characteristic morphologies in turbulences with different $\mathcal{M}_\text{turb}$. The supersonic case (1024M7-b0.1, top panels) exhibits density concentrations of dots and strings, which are filaments and sheets in 3D, as was previously discussed for hydrodynamic turbulence \citep[see, e.g.,][]{Kim2005} and MHD turbulence \citep[see, e.g.,][]{Lehmann2016}. The subsonic case (1024M0.5-b10, bottom panels), on the other hand, includes curves of discontinuities, which are surfaces of shocks with density jumps of mostly $\lesssim 2$.

Figure \ref{fig3} shows the 3D distributions of $M_\text{s}$ for fast and slow shocks, from the same data as in Figure \ref{fig2}. The images of $M_\text{s}$ reveal a few points. Note that while $M_\text{s} > 1$ for fast shocks, $M_\text{s}$ could be smaller than unity for slow shocks. First, the distribution of $M_\text{s}$ shows a clear correlation to that of $\bm{\nabla}\cdot\bm{v}$, as expected. Second, fast shock populations dominates over slow shocks for 1024M0.5-b10, a weak background magnetic field case. On the other hand, slow shocks are as frequent as fast shocks for 1024M7-b0.1, a strong magnetic field case. This trend is true, regardless of $\mathcal{M}_\text{turb}$, as further discussed in Section \ref{sec4.2}. Third, shock surfaces are not smooth nor homogeneous, but composed of shocks with different Mach numbers. Yet, fast shocks show organized structures of strings in 2D cuts and surfaces in 3D distributions. Slow shocks, on the other hands, display fragmented structures. {This could be partly because the surfaces of some slow shocks may be subject to a corrugation instability} \citep[see, e.g.,][]{Stone1995}, as noted above.

\subsection{Statistics of Shock Mach Numbers}\label{sec4.2}

The PDFs of $M_\text{s}$ for fast shocks (upper panels) and slow shocks (lower panels) in the cases of $\mathcal{M}_\text{turb}=0.5$, 1, and 7, for which both $1024^3$ and $512^3$ simulations are available, are shown in Figures \ref{fig4}. Only fast shocks with $M_\text{fa}\geq1.06$ and slow shocks with $M_\text{sl}\geq1.06$ and $c_\text{sl,1}/c_\text{s} \geq 0.3$ were included, as described in Section \ref{sec3.2}. The PDFs were obtained by averaging those from 16 and 26 data dumps for $1024^3$ and $512^3$ simulations, respectively, over the saturated state. They were normalized with $n^2_\text{g}$ to compensate the resolution effect. {The average values of $M_\text{s}$ for all models are given Table \ref{Table2}.}  Simulations of different resolutions produced reasonably converged results.

A noticeable feature is the ``flat'' parts with $M_\text{s} < 1$ for slow shocks in the lower panels of Figure \ref{fig4}. With $c_\text{sl,1} < c_\text{s}$, those shocks still have slow Mach numbers $M_\text{sl} > 1$. They substantially contribute to the population of slow shocks. Yet, these are the slow shocks affected by the difficulty in their identification (see Section \ref{sec3.2}); different values in the constraint of $c_\text{sl,1}/c_\text{s}$ result in differences in this part (see Appendix \ref{secA1}). There is almost no such population in the case of $\mathcal{M}_\text{turb} = 0.5$ and $\beta_0 = 0.1$, since weak perturbations propagating mostly perpendicular to strong background fields are hard to develop into shocks.

Another noticeable feature is the lack of fast shocks with $M_\text{s} \lesssim 2-3$ for $\mathcal{M}_\text{turb} \lesssim 1$ in the case of $\beta_0=0.1$, in the upper panels of Figures \ref{fig4}; with subsonic/transonic flows, the strong background magnetic field is only mildly perturbed, and hence $c_\text{fa,1} >$ a few $\times\ c_\text{s}$. In the cases of weak background magnetic fields or highly supersonic flows, magnetic field lines are easily tangled and turbulence becomes isotropic, and hence the PDFs follow the usual shape, which is described below.

Except the ``anomalies'' described above, the PDFs are characterized as follows. Most shocks are ``weak'' with low sonic Mach numbers, as previous shown \citep[see, e.g.,][]{Smith2000b,Porter2015,Lehmann2016}. Shocks with high Mach numbers are rare and mostly fast shocks. They show exponentially decreasing probability distributions. When the exponential part of high $M_\text{s}$ is fitted to
\begin{equation}
\frac{dN_\text{fa}}{dM_\text{s}} \propto \exp\left(-\frac{M_\text{s}}{M_{\text{cha},N}}\right), \label{eq21}
\end{equation}
the characteristic Mach number, $M_{\text{cha},N}$, increases with $\mathcal{M}_\text{turb}$, but decreases with $\beta_0$ although the dependence on $\beta_0$ is not strong. For instance, $M_{\text{cha},N} \simeq 0.1$ for 1024M0.5-b10 with $\mathcal{M}_\text{turb}=0.5$ and $\beta_0=10$; it was estimated to be $\sim 0.08$ for $\mathcal{M}_\text{turb}=0.5$ and $\beta_0=10^6$ in \citet{Porter2015}. For 1024M7-b0.1 with $\mathcal{M}_\text{turb}=7$ and $\beta_0=0.1$, on the other hand, $M_{\text{cha},N} \simeq 3$, indicating that shocks with high $M_\text{s}$ are quite common. For the cases listed in Table \ref{Table1}, the ratio $M_{\text{cha},N}/\mathcal{M}_\text{turb}$ is in a rather narrow range of $\sim 0.2 - 0.4$, except for 1024M0.5-b0.1 with $\mathcal{M}_\text{turb}=0.5$ and $\beta_0=0.1$, where fast shocks with high $M_\text{s}$ are rare, so their PDF is not very well defined.

Figure \ref{fig5} plots the total number of shock zones for fast and slow shocks and all together, $N_\text{fa}$, $N_\text{sl}$, and $N_\text{fa+sl}$, normalized to $n^2_\text{g}$, in all the cases considered (see also Table \ref{Table1} for the values). The trend is summarized as follows. First, $N_\text{fa+sl}$ increases with $\mathcal{M}_\text{turb}$ and $\beta$; that is, shocks are more common or shock surface area is larger if the turbulent flow velocity is larger and the background magnetic field is weaker. Second, fast shocks are more common than slow shocks for $\beta_0 = 10$. For $\beta_0 = 0.1$, on the other hand, slow shocks are as frequent as fast shocks, or even more frequent if $\mathcal{M}_\text{turb}$ is small.

The spatial frequency of shocks may be presented as the {\it mean distance between shock surfaces}, as was done in \citet{Ryu2003}.  For fast and slow shocks altogether, that is, with $N_\text{fa+sl}$ in the bottom panel of Figure \ref{fig5}, the mean distance is estimated to be, for instance, $\sim (1/2) L_\text{inj}$ for 1024M0.5-b10 and 1024M7-b0.1, while it is $\sim (1/10) L_\text{inj}$ for 1024M7-b10 with $\mathcal{M}_\text{turb}=7$ and $\beta_0=10$.

\subsection{Energy Dissipation at Shocks}\label{sec4.3}

Figures \ref{fig6} shows the distributions of the energy dissipation rate in Equation (\ref{eq18}), normalized to the energy injection rate in Equation (\ref{eq20}), as functions of $M_\text{s}$, for fast shocks (upper panels) and slow shocks (lower panels) in the cases of $\mathcal{M}_\text{turb}=0.5$, 1, and 7. They were obtained by averaging the data over the saturated state. Again, simulations of different resolutions produced reasonably converged results.

As for the PDFs of $M_\text{s}$, the energy dissipation is contributed mostly by shocks with low Mach numbers, although the distributions of the energy dissipation have the peaks at higher $M_\text{s}$ than the PDFs; the peak locates close to $M_\text{s} \sim 1$ for 1024M0.5-b10 with $\mathcal{M}_\text{turb}=0.5$ and $\beta_0=10$, while it occurs at $M_\text{s} \sim$ a few to several in the cases of $\mathcal{M}_\text{turb}=7$. At shocks with high $M_\text{s}$, the energy dissipation per unit area, on average, scales as $\propto M_\text{s}^3$ (see Equation (\ref{eq15})), but the area decreases exponentially with $M_\text{s}$. Hence, the energy dissipation rate also shows exponentially decreasing distributions of $M_\text{s}$. When the exponential part of high $M_\text{s}$ is fitted to
\begin{equation}
\frac{d\epsilon_\text{fa}}{dM_\text{s}} \propto \exp\left(-\frac{M_\text{s}}{M_{\text{cha},\epsilon}}\right), \label{eq22}
\end{equation}
the characteristic Mach number, $M_{\text{cha},\epsilon}$, is larger for larger $\mathcal{M}_\text{turb}$ and smaller $\beta_0$. For instance, $M_{\text{cha},\epsilon}$ is estimated to be $\sim0.14$ and $\sim3.5$ for 1024M0.5-b10 and 1024M7-b0.1, respectively. For the cases considered here, the ratio $M_{\text{cha},\epsilon}/\mathcal{M}_\text{turb}$ is in the range of $\sim 0.3 - 0.5$, except for 1024M0.5-b0.1.

In general, the amount of the energy dissipated at fast shocks is larger than that at slow shocks (see also Figure \ref{fig7}). The bottom panels of Figure \ref{fig6}, however, indicate that the energy dissipation at slow shocks with $M_\text{s} < 1$ is substantial, especially in the cases of low $\mathcal{M}_\text{turb}$. This is mostly because the MHD part inside the parenthesis in Equation (\ref{eq15}), i.e., the term involving $v_\text{A}$, is positive for slow shocks, while it is negative for fast shocks; the term accounts for the decrease of the magnetic energy across the slow shock. On the top of that, as {listed in Table \ref{Table2},} also previously shown by \citet{Lehmann2016}, slow shocks, on average, form in preshock conditions with stronger magnetic fields than fast shocks. However, for slow shocks in the range of $M_\text{s} < 1$, the population was estimated with large uncertainties, as described in Section \ref{sec4.2}, so should have been the energy dissipation.

Bearing the uncertainties in mind, Figure \ref{fig7} plots the fraction of the total energy dissipated at fast and slow shocks and all together, normalized to the injected energy, in all the cases considered (see also Table \ref{Table1} for the values). {\it The fraction decreases with increasing} $\mathcal{M}_\text{turb}$, except for low $\mathcal{M}_\text{turb}$ and low $\beta_0$, where the number of shocks is very small, so is the energy dissipation. The fraction of the total shock energy dissipation, $\epsilon_\text{fa+sl}/\epsilon_\text{inj}$, is in the range of $\sim0.1-0.6$. It is close to $\sim0.4-0.6$ in the cases of subsonic turbulences, $\mathcal{M}_\text{turb}=0.5$, with $\beta_0=1-10$. The fraction is smaller with $\sim0.1-0.25$ for highly supersonic turbulences, $\mathcal{M}_\text{turb}=7$. The dissipation at fast shocks is larger by at least a few times, and sometimes a few tens times, than that at slow shocks, partly because fast shocks have higher Mach numbers, $M_\text{s}$, but also because fast shocks are more abundant. An exception is the case of 1024M1-b1 with $\mathcal{M}_\text{turb}=1$ and $\beta_0=1$, for which the energy dissipations at both modes of shocks are comparable. However, this is perhaps the case where the uncertainties could have seriously affected the statistics, as noted above.

The rest of the turbulent energy should dissipate through the turbulent cascade. The fractions of the cascade dissipation (Equation (\ref{eq19})), normalized to the energy injection, $\epsilon_\text{cas}/\epsilon_\text{inj}$, are listed in Table \ref{Table1}. Again, they were obtained by averaging the data over the saturated state. The values are in the range of $\sim 0.5 - 1$. The total dissipation, $\epsilon_\text{fa+sl} + \epsilon_\text{cas}$, is, however, not always equal to the injection, $\epsilon_\text{inj}$. It is larger than $\epsilon_\text{inj}$ in some cases, but smaller in other cases. This should reflect uncertainties in $\epsilon_\text{cas}$; its estimation seems to be trustworthy only within a factor of two or so, due to the fact that in compressible MHD turbulences, the spectral slopes are not always equal to the Kolmogorov value. Yet, with the values of $\epsilon_\text{fa+sl}/\epsilon_\text{inj}$ and $\epsilon_\text{cas}/\epsilon_\text{inj}$ in Table \ref{Table1}, we conclude that in most cases, the energy dissipation through turbulent cascade would be larger than or at least comparable to the energy dissipation at shocks. In the cases of highly supersonic turbulences, $\epsilon_\text{cas}$ would be a few to several time larger than $\epsilon_\text{fa+sl}$; in the cases of subsonic turbulences, $\epsilon_\text{cas}$ would be comparable to $\epsilon_\text{fa+sl}$.

As described in the Introduction, previous works estimated the fraction of the energy dissipation through turbulent cascade, $t_\text{diss}/t_\text{cross}$ or $\epsilon_\text{cas}/\epsilon_\text{inj}$ in our notation. It was, for instance, {$\sim 0.46 - 0.69$} from simulations of supersonic turbulences \citep{Stone1998}, and $\sim 1/3$ for turbulence in the Perseus MC \citep{Pon2014} and $\sim 0.65$ or 0.94 for the Taurus MC \citep{Larson2015}. These values are in a rough agreement with our results.

\section{Summary and Discussion}\label{sec5}

In this paper, we studied shock waves in compressible, MHD turbulences with Mach numbers $\mathcal{M}_\text{turb}=0.5-7$ and initial magnetic fields of plasma beta $\beta_0=0.1 - 10$, using a set of ``isothermal'' simulations with up to $1024^3$ grid zones. The ranges of the parameters were chosen to cover turbulences in the ISM and ICM.  {Turbulence was driven by ``solenoidal'' forcing, leaving the examination of the dependence on the nature of forcing as a follow-up work.} We separately identified fast and slow shock populations; slow shocks are harder to be reliably identified, partly because their surfaces are subject to a corrugation instability \citep[e.g.,][]{Stone1995} and also because some of them have sonic Mach numbers $M_\text{s} < 1$. We then obtained the PDFs of $M_\text{s}$ and calculated the dissipation of turbulent energy at shocks. In order to minimize confusions between weak shocks and waves, only fast shocks with fast Mach numbers $M_\text{fa} \geq 1.06$ and slow shocks with slow Mach numbers $M_\text{sl} \geq 1.06$ and slow wave speeds $c_\text{sl,1} \geq 0.3 c_\text{s}$ were considered for the statistics presented. Our main findings are summarized as follows.

1. Most shocks form with low $M_\text{s}$. Strong shocks with high $M_\text{s}$, which are mostly fast shocks, are rare and follow exponentially decreasing probability distributions, $\propto \exp(-{M_\text{s}}/{M_{\text{cha},N}})$. The characteristic Mach number, $M_{\text{cha},N}$, is larger for larger $\mathcal{M}_\text{turb}$, indicating that shocks with high $M_\text{s}$ are more common if turbulent flows have higher speeds, as expected. In addition, $M_{\text{cha},N}$ is larger for smaller $\beta_0$, although the dependence on $\beta_0$ is not strong. The ratio $M_{\text{cha},N}/\mathcal{M}_\text{turb}$ is in the range of $\sim 0.2 - 0.6$ for the cases studied in this paper.

2. More shocks are induced, if turbulent flows have higher speeds and the background magnetic field is weaker. Fast shocks are more common than slow shocks, if $\beta_0 \gg 1$, i.e., in weak field cases. Slow shocks, on the other hand, are as frequent as, or even more common than, fast shocks, if $\beta_0 \lesssim 1$, i.e., in strong field cases. The mean distance between shock surfaces is estimated to be $\sim 1/2$ of the injection scale, $L_\text{inj}$, for typical turbulences expected in the ICM and MCs. It is smaller in the cases with highly supersonic turbulences in weak background magnetic fields.

3. The energy dissipation is mostly due to shocks with low $M_\text{s}$. The peak of the dissipation locates close to $M_\text{s} \sim 1$ for subsonic turbulences expected in the ICM, while it is at $M_\text{s} \sim$ a few to several for highly supersonic turbulences in MCs. The energy dissipation at shocks with high $M_\text{s}$ follows exponentially decreasing distributions, $\propto \exp(-{M_\text{s}}/{M_{\text{cha},\epsilon}})$; this is because while the dissipation per unit area of shock surfaces, on average, increases as $\propto M_\text{s}^3$, the shock population exponentially decreases with $M_\text{s}$. The ratio $M_{\text{cha},\epsilon}/\mathcal{M}_\text{turb}$ is in the range of $\sim 0.3 - 0.5$ for the cases studied in this paper.

4. The energy dissipation is attributed mostly to fast shocks, partly because fast shocks are stronger than slow shocks, but also because fast shocks are more common in most cases. {\it The fraction of the turbulent energy dissipated at shocks}, $\epsilon_\text{fa+sl}/\epsilon_\text{inj}$ (the energy dissipation at both fast and slow shocks, normalized to the energy injection), {\it decreases with increasing} $\mathcal{M}_\text{turb}$, ranging from $\sim 0.1-0.25$ for highly supersonic turbulences with $\mathcal{M}_\text{turb}=7$ to $\sim 0.4-0.6$ for subsonic turbulences with $\mathcal{M}_\text{turb}=0.5$. Note that both $\epsilon_\text{fa+sl}$ and $\epsilon_\text{inj}$ increase with $\mathcal{M}_\text{turb}$; however, $\epsilon_\text{inj}$ increases faster than $\epsilon_\text{fa+sl}$.  The rest of the turbulent energy should dissipate through the turbulent cascade.

The statistics presented in this paper should be applicable within the context of the models considered here. For turbulences with different characteristics (different forcings, different equations of state (EoSs), and etc), quantitative estimates would be different, although we still expect that more shocks form if turbulent flows have higher speeds, fast shock populations dominate over slow shocks if the background magnetic field is weak, and etc.

Finally, our work would have implications for physical processes in turbulent ISM and ICM. For instance, the dissipation of turbulent energy at shocks may have consequences on observed shock structures and emission line spectra. We leave the investigation of those as future works.

\acknowledgments
{The authors thank the anonymous referee for the thorough review and constructive suggestions that lead to an improvement of the paper.} J.P. was supported by the National Research Foundation of Korea through grant 2016R1A5A1013277. D.R. was supported by the National Research Foundation of Korea through grant 2017R1A2A1A05071429.

\appendix

\section{Dependence on Shock Identification Parameters}\label{secA1}

\renewcommand\thefigure{\thesection.\arabic{figure}}   
\setcounter{figure}{0}

\begin{figure*}
\begin{center}
\vskip 0.3cm
\hskip -0.6cm
\includegraphics[width=1.02\textwidth]{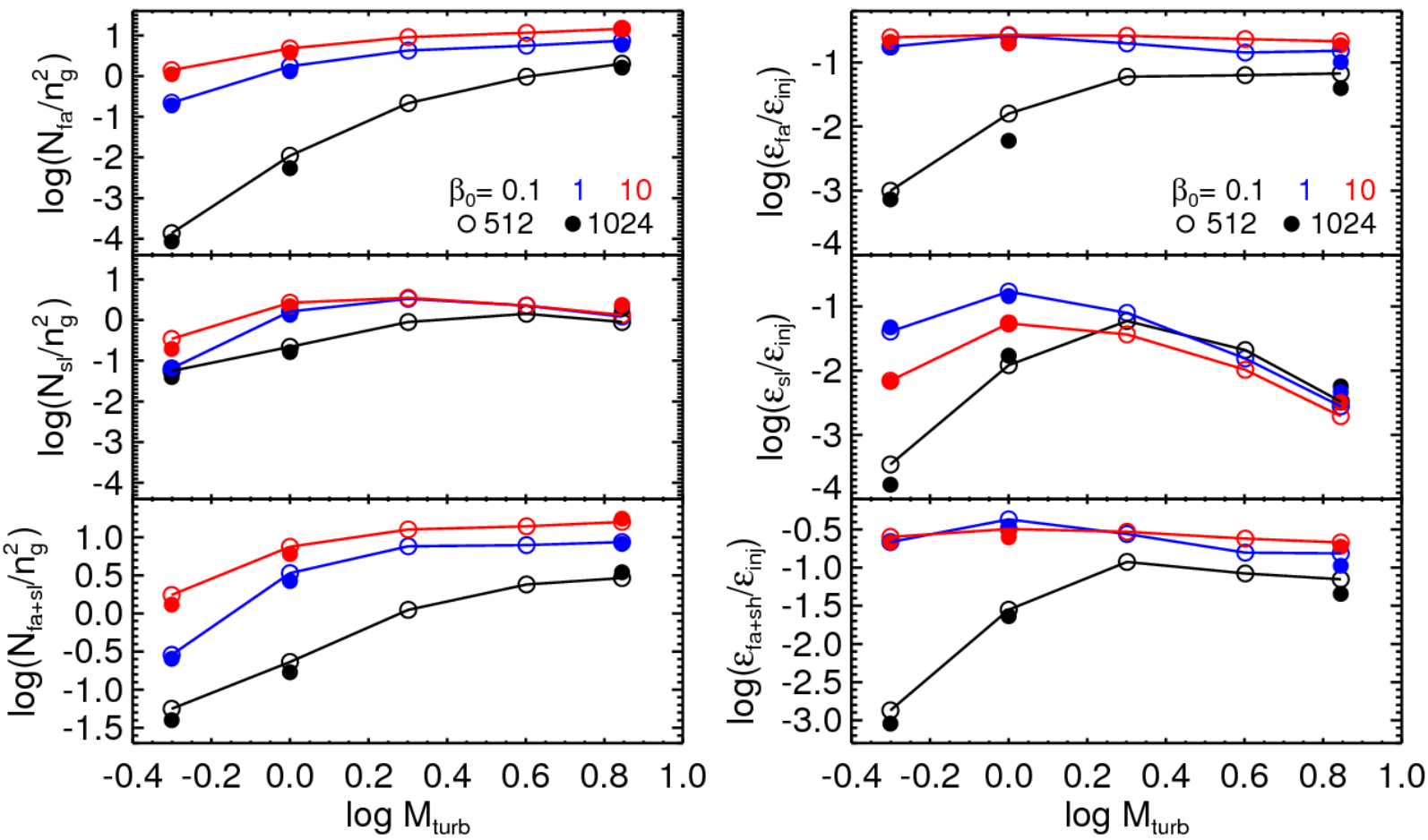}
\vskip -0.2cm
\caption{Statistics for shocks with $M_\text{fa,sl} \geq 1.1$. Left panels: number of the grid zones identified as fast (top) and slow (middle) shocks and their sum (bottom), normalized to $n_\text{g}^2$, as a function of $\mathcal{M}_\text{turb}$. Right panels: fraction of the energy dissipation at fast (top) and slow (middle) shocks and their sum (bottom) as a function of $\mathcal{M}_\text{turb}$.} \label{figa1}
\end{center} 
\end{figure*}

\begin{figure*}
\begin{center}
\vskip 0.3cm
\hskip -0.6cm
\includegraphics[width=1.02\textwidth]{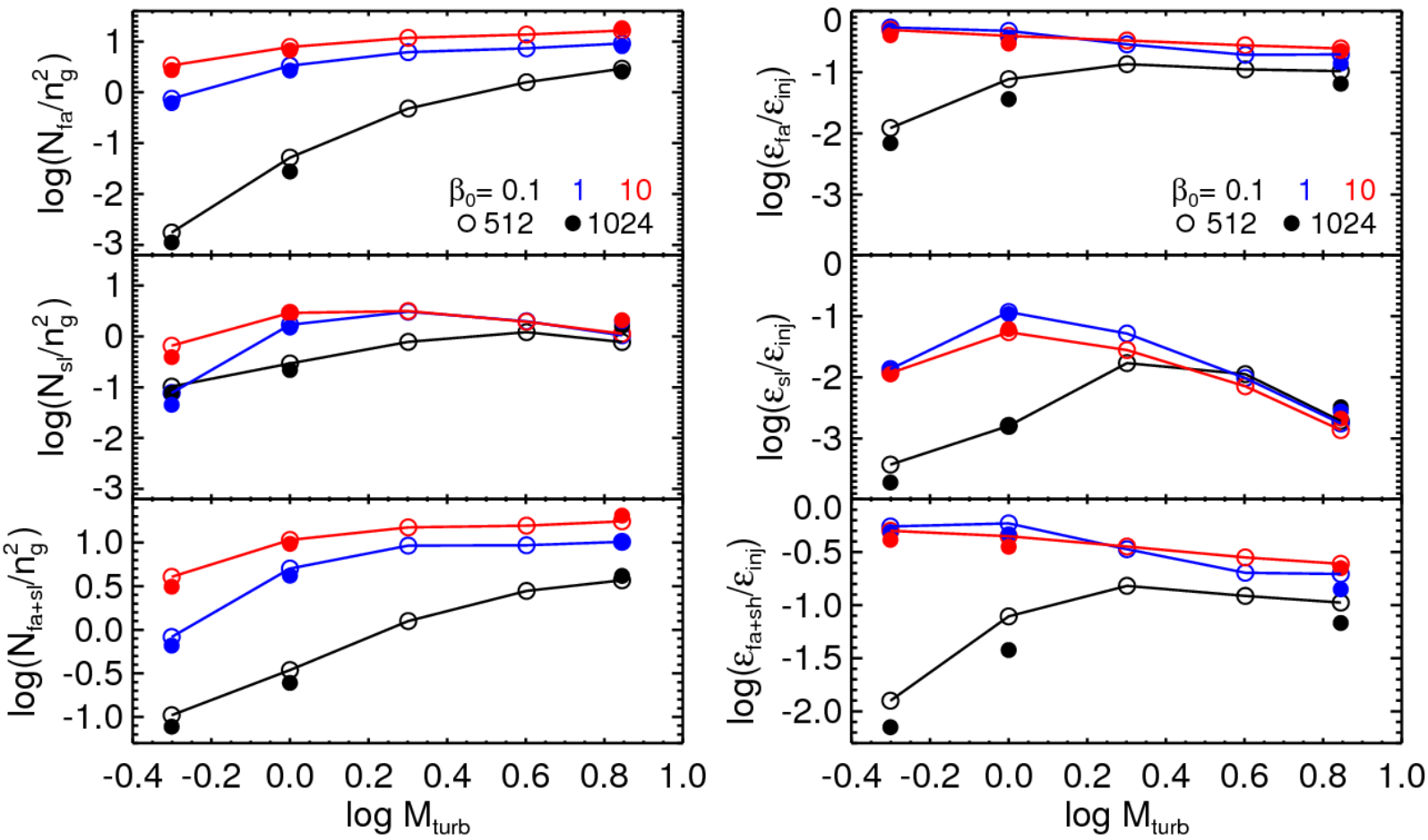}
\vskip -0.2cm
\caption{Statistics for shocks with $c_\text{sl,1}/c_\text{s} \geq 0.4$. Left panels: number of the grid zones identified as fast (top) and slow (middle) shocks and their sum (bottom), normalized to $n_\text{g}^2$, as a function of $\mathcal{M}_\text{turb}$. Right panels: fraction of the energy dissipation at fast (top) and slow (middle) shocks and their sum (bottom) as a function of $\mathcal{M}_\text{turb}$.} \label{figa2} 
\end{center}
\end{figure*}

The statistics present in Section \ref{sec4} inevitably depend on the parameters employed in the identification of shocks. We chose $\max(\rho_{i+1}/\rho_{i-1},\rho_{i-1}/\rho_{i+1}) \geq 1.03^2$, $M_\text{fa}\geq 1.06 $, $M_\text{sl} \geq 1.06$, and $c_\text{sl,1} \geq 0.3 \times c_\text{s}$, after a number of experiments with different values. Smaller lower bounds result in more confusions with waves, while larger values exclude some of shocks. Here, we present the statistics for a couple of different sets of parameter values to demonstrate the dependence of our results on them.

Figure \ref{figa1} shows the total number of shock zones normalized to $n^2_\text{g}$ (left panels) and the turbulent energy dissipation at shocks normalized to the energy injection (right panels), obtained with $M_\text{fa}\geq 1.1$ and $M_\text{sl} \geq 1.1$, keeping the minimum value of $c_\text{sl,1}$ same; $\max(\rho_{i+1}/\rho_{i-1},\rho_{i-1}/\rho_{i+1}) \geq 1.05^2$ was used, but once the minimum value of the square-root of it is smaller than those of $M_\text{fa}$ and $M_\text{sl}$, the results are insensitive to it. The trends still hold; the number of shock zones, $N_\text{fa+sl}/n_\text{g}^2$, is larger if turbulence has larger $\mathcal{M}_\text{turb}$ and the background magnetic field is weaker, and the fraction of the energy dissipation at shocks, $\epsilon_\text{fa+sl}/\epsilon_\text{inj}$, decreases with increasing $\mathcal{M}_\text{turb}$, except for the cases of low $\mathcal{M}_\text{turb}$ and low $\beta_0$. However, $N_\text{fa+sl}/n_\text{g}^2$ and $\epsilon_\text{fa+sl}/\epsilon_\text{inj}$ are smaller, as expected. The decrements are larger for smaller $\mathcal{M}_\text{turb}$. For instance, $\epsilon_\text{fa+sl}/\epsilon_\text{inj}$ is close to $\sim0.2-0.25$ for $\mathcal{M}_\text{turb}=0.5$ and $\beta_0=1-10$, while it is in the range of $\sim0.05-0.2$ for $\mathcal{M}_\text{turb}=7$.

Figure \ref{figa2} shows the total number of shock zones and the energy dissipation at shocks, obtained with $c_\text{sl,1} \geq 0.4 \times c_\text{s}$, keeping the lower bounds of $M_\text{fa}$ and $M_\text{sl}$ same. Obviously, the increase in the minimum value of $c_\text{sl,1}$ results in the decrease in slow shock populations, without affecting fast shock populations. But $N_\text{fa+sl}/n_\text{g}^2$ and $\epsilon_\text{fa+sl}/\epsilon_\text{inj}$ are not much affected, since they are contributed mostly from fast shock populations.

\end{document}